\documentclass[submission,copyright,creativecommons]{eptcs}
\providecommand{\event}{ICLP DC 2019} 

\usepackage{breakurl}             
\usepackage{underscore}  

\newcommand{\ocircle}{\odot}
\newcommand{\C}{\mathcal{C}}
\usepackage{amssymb}
\usepackage{amsthm}
\usepackage{amsmath}
\usepackage{listings}
\usepackage{multirow}

\theoremstyle{definition}
\newtheorem{definition}{Definition}[section]

\theoremstyle{plain}
\newtheorem{theorem}{Theorem}[section]

\usepackage{tikz}
\usetikzlibrary{positioning,shapes.arrows,shapes.misc,calc,automata,fit}

\usepackage{url}
\title{Distributed Answer Set Coloring: Stable Models Computation via Graph Coloring} 

\def\titlerunning{Distributed Computation of Answer Sets}

\author{Marco De Bortoli \institute{Institute for Software Technology, Graz University of Technology, Graz, Austria} \institute{Dept DMIF, University of Udine, Udine, Italy} \institute{Dept CS, New Mexico State University, Las Cruces, NM, USA}  \email{mbortoli@ist.tugraz.at}}

\def\authorrunning{Marco De Bortoli}

\begin{document}

\maketitle

\begin{abstract}
Answer Set Programming (ASP) is a famous logic language for knowledge representation, which has been really successful in the last years, as witnessed by the great interest into the development of efficient solvers for ASP. 
Yet, the great request of resources for certain types of problems, as the planning ones, still constitutes a big limitation for problem solving. Particularly, in the case the program is grounded before the resolving phase, an exponential blow up of the grounding can generate a huge ground file, infeasible for single machines with limited resources, thus preventing even the discovering of a single non-optimal solution. To address this problem, in this paper we present a distributed approach to ASP solving, exploiting distributed computation benefits in order to overcome the just explained limitations. The here presented tool, which is called Distributed Answer Set Coloring (DASC), is a pure solver based on the well-known Graph Coloring algorithm. 
DASC is part of a bigger project aiming to bring logic programming into a distributed system, started in 2017 by Federico Igne with mASPreduce and continued in 2018 by Pietro Totis with a distributed grounder.
In this paper we present a low level implementation of the Graph Coloring algorithm, via the Boost and MPI libraries for C++. Finally, we provide a few results of the very first working version of our tool, at the moment without any strong optimization or heuristic.

\end{abstract}

\begin{footnotesize}
I want to thank Fabio Tardivo, Agostino Dovier and Enrico Pontelli for their support during the development of the tool.
\end{footnotesize}

\section{Introduction and problem description}
The Answer Set Programming (ASP) language has become very popular in the last years thanks to the availability of more and more efficient solvers (e.g., Clingo  \cite{DBLP:journals/corr/GebserKKS14}
and DLV \cite{DBLP:journals/ki/AdrianACCDFFLMP18}).
It is based on the stable model semantics from Gelfond and Lifschitz \cite{Gelfond1990},
introduced to resemble the human reasoning process; together with its simple syntax, this makes ASP a very intuitive language to be used.  Like most logic languages, the ASP solving process is split into two phases: the \textit{grounding}, namely the transformation of the normal program into a so-called ground program, which is the equivalent propositional logic program where each rule is instantiated over the domain of its variables.
The second phase consists in the real solving process, which alternates non-deterministic guesses and deterministic propagation to find the solutions, starting from the ground program.
As described, e.g.,  in \cite{DBLP:journals/fuin/PaluDPR09}, ASP has some important weakness when dealing with real-world complex problems, like planning \cite{DBLP:journals/jetai/DovierFP09,Son2007}, which generates huge ground programs. The grounding phase is in fact a strong limitation when dealing with problems that generate a great amount of rules, especially if it is an in-memory computation.
This kind of programs leads to two issues, one regarding the grounding itself and one regarding the computation of its stable models, both limited by the amount of resources of the machine. Even if in literature there is a fair interest towards the parallelization of stable models computation, the single-machine multithreading applied to this field still has the memory limitation issue.
To address this problem, we present in this paper a distributed ASP solver, called Distributed Answer Set Coloring (DASC), which automatically splits a ground program over a network, using the resources of a single computational node to process only a portion of the original program. To accomplish this, we use the Graph Coloring algorithm, which represents the program as a graph, and its stable models as different colorings of its vertices.\\
DASC is not the first attempt of this sort: it was born with the purpose of lowering the implementation level of the mASPreduce solver, a tool developed by Federico Igne for his Master thesis \cite{ignethesis}, in order to address its performance. mASPreduce is indeed developed with the distribution framework Spark from Apache, which, although it is a very powerful and expressive framework for distributed programming, gives to the user very low control over the communication flow, and it is the real performance killer during such kind of distributed computations.\\
We handled this communication control problem by developing DASC with C++, using the MPI library for messages handling and the Parallel Boost Graph Library to represent the distributed graph to color.\\
This research summary is organised as follows. In Section 2, we give some basic notations about ASP semantics and the Graph Coloring algorithm, and we see some related work in literature. Then we present the main details of our tool in Section 3. Some experimental results and comparison between DASC, mASPreduce and the state-of-the-art Clingo solver are reported in Section 4. The reader can find our conclusions and future work in Section 5.

\section{Background and Related Work}

\subsection{Answer Set Programming} 
We provide  a quick review of the basic concepts of \emph{Answer Set Programming (ASP)}. We assume familiarity with logic programming.\\

A \emph{normal rule} $r$ is of the form
\begin{equation}
h \leftarrow a_1,\dots,a_n, \mathtt{not} \ b_1,\dots, \mathtt{not} \ b_m.
\end{equation}
where $0 \le n,m$, $h$ is a positive literal, for $1 \le i \le n$, $a_i$ is a positive literal, and for $1 \le j \le m$, $\mathtt{not} \ b_j$ is a negative literal. A rule where $m = n = 0$ is called
a \emph{fact}, and a rule where  $h = \bot$ (representing false) is called a \emph{constraint}. A \emph{normal logic program} $\Pi$ is a finite set of normal rules and $\textsc{Atoms}(\Pi)$ is the set of all atoms of the alphabet occurring in $\Pi$.
Given a normal rule $r$ of a program $\Pi$, we define $head(r) = h$,
$body^+(r) = \{a_1,\dots,a_n\}$, $body^-(r) = \{b_1,\dots,b_m\}$ and $body(r) = \{a_1,\dots,a_n, \mathtt{not} \ b_1,\dots, \mathtt{not} \ b_m\}$.
Given a program $\Pi$, we can write $head(\Pi) = \{head(r) \mid r \in \Pi\}$ (similar for $body$, $body^+$ and $body^-$).

We call a program $\Pi$ with $body^-(\Pi) = \emptyset$ and without constraints a \emph{definite} logic program.
A definite logic program always admits a minimum Herbrand model, denoted by $Cn(\Pi)$.

We say that a set $X$ of atoms satisfies a rule $r$ (in symbols $X \vDash r$) if $head(r) \in X$ whenever $body^+(r) \subseteq X$ and $body^-(r) \cap X = \emptyset$, and that $X$
satisfies a program $\Pi$ if for all $r \in \Pi$ $(X \vDash r)$.

Given a program $\Pi$ and a set $X \subseteq \textsc{Atoms}(\Pi)$, we define the \emph{reduct} $\Pi^X$ of $\Pi$ w.r.t.\ $X$ as $  \Pi^X = \{head(r) \leftarrow body^+(r) \mid r \in \Pi, body^-(r) \cap X = \emptyset\}$. $X$ is an \emph{answer set} of $\Pi$ iff $Cn(\Pi^X) = X$. $AS(\Pi)$ denotes the set of all answer sets of a program $\Pi$.

The operator $Cn(\cdot)$ can also be characterized as the least fixpoint of the  \emph{immediate consequence operator}  $T_\Pi(X) = \{head(r) \mid r \in \Pi, body^+(r) \subseteq X\}$.
Iterated applications of $T_\Pi$ can be defined as
$T_\Pi^0(X) = X$ and $T_\Pi^{i+1}(X) = T_\Pi(T_\Pi^i(X))$, for $i \ge 1$.
For a definite logic program $\Pi$, it can be proven that $Cn(\Pi) = \bigcup_{i \ge 0} T_\Pi^i(\emptyset)$.
Therefore, $X$ is an answer set of $P$ iff
$\bigcup_{i \ge 0} T_{\Pi^X}^i(\emptyset) = X$.

\begin{definition}[Generating Rules of an answer set] 
Given a set of atoms $X$ from a program $\Pi$, the set $\mathcal{R}_\Pi(X)$ of generating rules is given by
\[ R_\Pi(X) = \{ r \in\Pi \: | \: body^+(r) \subseteq  X, body^-(r) \cap X = \emptyset\}.  \]
\end{definition}
$R_\Pi(X)$ is uniquely identified by an answer set $X$.
Given a program $\Pi$, a set $X \subseteq \textsc{Atoms}(\Pi)$ is an \emph{answer set} of $\Pi$ iff $Cn(R_\Pi(X)^\emptyset) = X$ \cite{DBLP:journals/tplp/KonczakLS06}---let us observe that $R_\Pi(X)^\emptyset$ is the reduct of $R_\Pi(X)$ w.r.t.\ the empty set of atoms.

The semantics discussed above assumes that the atoms in the program are ground, i.e., they do not contain any variables.
Intuitively, an atom/rule containing variables is a shorthand for the set of all possible ground instances obtained by consistently replacing each variable with all possible elements of the Herbrand universe. The process of rewriting a rule with variables into the equivalent set of rules without variables is referred to as  \emph{grounding}.

\subsection{The Graph Coloring Algorithm}
We briefly present the Coloring algorithm for the computation of answer sets \cite{DBLP:journals/tplp/KonczakLS06}, used by both mASPreduce and DASC solvers.

A \emph{labeled graph} is a pair $(G,\ell)$ where $G = (V,E)$ is a directed graph and $\ell : E \rightarrow \mathcal{L}$ is a mapping from edges to the set $\mathcal{L} = \{0,1\}$ of labels 
(intuitively $0$ will represent a positive dependency and $1$ a negative dependency).
$(G,\ell)$ can be represented by the triple  $(V,E_0,E_1)$, where $E_i = \{  e \in E \mid \ell(e) = i\}$ for $i = 0,1$.
Given a labeled graph $G = (V,E_0,E_1)$, an $i$\emph{--subgraph} of $G$ for $i = 0,1$ is a
subgraph of the graph $G_i = (V,E_i)$---i.e. a graph $G' = (W,F)$ s.t.\ $W \subseteq V$, and $F \subseteq E_i \cap (W^2)$.
If $x,y \in V$, an $i$\emph{--path} is a path from $x$ to $y$ in the graph  $G_i$.

Let $\Pi$ be a ground logic program; its \emph{rule dependency graph ($\mathcal{RDG}$)} $\Gamma_\Pi = (\Pi, E_0,E_1)$  is a labeled graph where nodes are the program rules and
    \begin{align*}
        E_0 & = \{(r,r') \mid r,r' \in \Pi, head(r) \in body^+(r')\}\\%
        E_1 & = \{(r,r') \mid r,r' \in \Pi, head(r) \in body^-(r')\}
    \end{align*}
A \emph{(partial/total) coloring} of $\Gamma_\Pi$ is a partial/total mapping
$C : \Pi \rightarrow \{\oplus,\ominus\}$, where $\oplus$ and $\ominus$ are two colors.
We will denote $C_\oplus = \{r \mid r \in \Pi, C(r) = \oplus\}$ and $C_\ominus = \{r \mid r \in \Pi, C(r) = \ominus\}$, and a (partial) coloring as $(C_\oplus,C_\ominus)$.
Let $\mathbb{C}_\Pi$ be the set of all (partial) colorings, and define a partial order over $\mathbb{C}_\Pi$ as follows: let $C,C'$ be partial coloring of $\Gamma_\Pi$. We say that $C \sqsubseteq C'$ iff $C_\oplus \subseteq C'_\oplus$ and $C_\ominus \subseteq C'_\ominus$. The empty coloring $(\emptyset,\emptyset)$ is the \emph{bottom} of the partial order $\mathbb{C}_\Pi$.
Colors represent \emph{enabling} ($\oplus$) and \emph{disabling} ($\ominus$) of rules.  Intuitively, we are interested in finding all possible sets of \emph{generating rules},  leading us to all the possible answer sets of a logic program.

Let $\Pi$ be a logic program, and let $\Gamma_\Pi$ be the corresponding $\mathcal{RDG}$.
We define the notion of  \emph{admissible coloring}  as follows:
if  $X \in AS(\Pi)$, then $C = (R_\Pi(X),\Pi \setminus R_\Pi(X))$ is an \emph{admissible coloring}  of $\Gamma_\Pi$ (i.e.,  all the rules whose bodies are satisfied by $X$ are colored positively, and the other rules negatively) such that $head(C_\oplus) = X$  \cite{DBLP:journals/tplp/KonczakLS06}.
We denote by $AC(\Pi)$ the set of all admissible colorings of $\Gamma_\Pi$.

By definition, admissible colorings are total and correspond one-to-one with answer sets. As shown in \cite{DBLP:journals/tplp/KonczakLS06}, for computing them we have to visit the space of partial colorings. Of course we are interested in partial colorings that will lead us to a total admissible coloring.

Let $\Pi$ be a program and $C$ a coloring of $\Gamma_{\Pi} = (\Pi,E_0,E_1)$.
For $r \in \Pi$:
    \begin{itemize}
        \item $r$ is \emph{supported} in $(\Gamma_{\Pi} ,C)$, if $body^+(r) \subseteq \{head(r') \mid (r',r) \in E_0,r' \in C_\oplus\}$;
        \item $r$ is \emph{unsupported} in $(\Gamma_{\Pi} ,C)$, if there is $q \in  body^+(r)$ s.t.\ $\{r' \mid (r',r) \in E_0, head(r') = q\} \subseteq C_\ominus$;
        \item $r$ is \emph{blocked} in $(\Gamma_{\Pi} ,C)$, if there exists $r' \in C_\oplus$ s.t.\ $(r',r) \in E_1$;
        \item $r$ is \emph{unblocked} in $(\Gamma_{\Pi} ,C)$, if $r' \in C_\ominus$ for all $(r',r) \in E_1$.
    \end{itemize}

We also define the sets of  supported $S(\Gamma,C)$,
unsupported $\bar{S}(\Gamma,C)$, blocked  $B(\Gamma,C)$, and unblocked  $\bar{B}(\Gamma,C) $ rules.
By definition,
${S}(\Gamma,C) \cap \bar{S}(\Gamma,C) = \emptyset$ and
${B}(\Gamma,C) \cap \bar{B}(\Gamma,C) = \emptyset$.
With $C$ a total coloring, a rule is unsupported or unblocked iff it is not supported or blocked, respectively. This is not true, in general, for partial colorings.

The above defined notions can be used to define an operational semantics  to compute the stable models of a logic program. We will only give an overview of the characterization implemented in our solver. For a deeper analysis of several other operational characterizations, we refer the reader to \cite{DBLP:journals/tplp/KonczakLS06}.

Let $\Gamma$ be the $\mathcal{RDG}$ of a logic program $\Pi$ and $C$ be a partial coloring of $\Gamma$. The coloring operator $\mathcal{D}_\Gamma^\ocircle: \mathbb{C} \rightarrow \mathbb{C}$, where $\ocircle \in \{\oplus, \ominus\}$, is defined as follows:
\begin{enumerate}
  \item $\mathcal{D}_\Gamma^\oplus = (C_\oplus \cup \{r\}, C_\ominus)$ for some $r \in S(\Gamma,C) \setminus (C_\oplus \cup C_\ominus)$;
  \item $\mathcal{D}_\Gamma^\ominus = (C_\oplus, C_\ominus \cup \{r\})$ for some $r \in S(\Gamma,C) \setminus (C_\oplus \cup C_\ominus)$.
\end{enumerate}

Operator $\mathcal{D}_\Gamma^\ocircle$ will be used to encode a branching path in the visit of the coloring tree, in fact, representing a non-deterministic choice (restricting our choice to the supported rules). Since \emph{support} is a local property of a node (it only depends on information coming from the neighborhood), the coloring operator can be efficiently applied.

Let $\Gamma$ be the $\mathcal{RDG}$ of a logic program $\Pi$ and $C$ be a (partial) coloring of $\Gamma$. Let us define the operators $\mathcal{P}_\Gamma,\mathcal{T}_\Gamma ,\mathcal{V}_\Gamma : \mathbb{C} \rightarrow \mathbb{C}$ as follows
$$\begin{array}{rcl}
    \mathcal{P}_\Gamma(C)  & = & (C_{\oplus} \cup (S(\Gamma,C) \cap \bar{B}(\Gamma,C)),C_{\ominus} \cup(\bar{S}(\Gamma,C) \cup B(\Gamma,C)))\\
      \mathcal{T}_\Gamma(C) & = & (C_\oplus \cup (S(\Gamma,C) \setminus C_\ominus), C_\ominus)\\
    \mathcal{V}_\Gamma(C) &= &(C_\oplus, \Pi \setminus V) \end{array}$$
where $V = \mathcal{T}_\Gamma^*(C_\oplus)$ and $\mathcal{T}_\Gamma^*(C)$ is the $\sqsubseteq$-smallest coloring containing $C$ and closed under $\mathcal{T}_\Gamma$. A coloring $C$ is closed under the operator $\mathbin{op}$ if $C = \mathbin{op}(C)$.
Finally, let
$(\mathcal{PV})_\Gamma^*(C)$ be the $\sqsubseteq$-smallest coloring containing $C$ and being closed under $\mathcal{P}_\Gamma$ and $\mathcal{V}_\Gamma$.

\begin{theorem}
[Operational Answer Set Characterization, III]\label{teo:oascIII}
    Let $\Gamma$ be the $\mathcal{RDG}$ of a logic program $\Pi$ and let $C$ be a total coloring of $\Gamma$.
Then, $C$ is an admissible coloring of $\Gamma$ iff there exists a \emph{coloring sequence} $C^0,C^1,\dots, C^n$ such that:
(1) $C^0 = (\mathcal{PV})_\Gamma^*((\emptyset,\emptyset))$,
(2) $C^{i+1} = (\mathcal{PV})_\Gamma^*(\mathcal{D}_\Gamma^\ocircle(C^i))$ for some $\ocircle \in \{\oplus,\ominus\}$ and $0 \le i < n$,
(3) $C^n = C$.
\end{theorem}

Given an admissible coloring $C$, $head(C_\oplus)$ returns its corresponding answer set. The proof of the above theorem can be found in \cite{DBLP:journals/tplp/KonczakLS06} and a general introduction of so-called ASP computation
is given in~\cite{DBLP:conf/iclp/LiuPST07}.

\subsection{A survey on parallel solving}
In this section we give a brief overview of existing parallelization techniques for ASP solving. For more details, we refer the reader to \cite{DBLP:books/sp/18/DovierFP18}.\\
\subsubsection{Parallel Grounding}
We are going to start by presenting a multi-level parallel approach for grounding, made up of three phases: the Component Level Parallelism, the Rule Level Parallelism and the Single-Rule Level Parallelism \cite{DBLP:journals/jal/CalimeriPR08}.\\
\newline
In the Component Level Parallelism, the dependency graph of the ASP program is split into several partitions (or components), according to the Strongly Connected Components (SCC) of the graph.\\
Then, we can make a grounder work on each component separately, in order to process all of them in parallel.\\
In the second phase, the rules of each module are grounded in parallel, following some guidelines in order to not do invalidate the result. In fact, the rules are split into two groups, the exit rules and the recursive rules, and the former are instantiated first. For more details see the original paper \cite{DBLP:books/sp/18/DovierFP18}.\\
Then, the third phase consists of splitting a rule body atom, selected by a heuristic procedure, and partitioning its extension between several threads.
\subsubsection{Parallel Solving}
Almost all the NP-problem solving techniques, including ASP solvers, are based on a search process. It can be seen as a tree (called search tree), in which each internal node represents a non-deterministic choice, each leaf is either a solution or a inconsistency, and the edges between different nodes are the deterministic propagation part which leads from a non-deterministic guess to the following one.\\
The general idea to parallelize this process is to assign to different threads (or computational nodes) the processing of the several subtrees generated along the search \cite{DBLP:conf/asp/PontelliE01,DBLP:conf/asp/FinkelMMT01}.\\
However, this approach has two main problems:
\begin{itemize}
\item
the knowledge/information collected till the moment a subtree $T$ is split into more parts (for instance, a subtree for each of $T$'s children) has to be replicated, possibly causing a memory blow up;
\item
to improve performance, solvers use heuristics which collect information from a specific branch in order to reuse it on another branches. This requires communication between the different parallel processes.
\end{itemize}
To address the latter, various techniques have been developed, divided into two groups: task sharing and scheduling. For more details, we refer the reader to the original paper \cite{DBLP:books/sp/18/DovierFP18}.

\subsubsection{Related Work}
Our solver is part of a bigger project aiming to bring logic programming into a distributed system, started in 2017 by Federico Igne with mASPreduce \cite{ignethesis,DBLP:conf/iclp/IgneDP18} and continued in 2018 by Pietro Totis with the distributed grounder STRASP \cite{totisthesis}.\\
Both of them are implemented in Scala, with the distribution framework Spark from Apache. The main difference between them lies in their purposes:
\begin{itemize}
\item
mASPreduce is a pure solver, and it is a high level implementation of the Graph Coloring algorithm. Spark provides support for automatic distribution and parallelization via the MapReduce technique \cite{Dean2008}: as a consequence, all the Graph Coloring operators are implemented as MapReduce routines. Although the operators seem pretty suitable to be encoded in this way, this leads to a network overhead, seriously affecting the performance;
\item
STRASP is a multi-purpose distributed tool: it can be used as a grounder for any kind of ASP programs, or as a solver for definite and stratified programs, like most of the grounders.\\
It makes use of the two first levels of parallelism during the grounding phase: the Component Level Parallelism and the Rule Level Parallelism.\\
Unfortunately, it suffers of the same performance issues as the previous system, because of the underlying framework.
\end{itemize}

In view of this, we chose for a different strategy: since communication handling is as important as the solving algorithm in a distributed environment, we developed from scratch a new implementation for the Graph Coloring algorithm, using low level languages and frameworks, such as C++, the Boost libraries and the MPI library for communication. This has led to a considerable performance improvement, as the reader can see in Section 4.

\section{Distributed Answer Set Coloring}
The DASC solver has been developed
with the purpose of improving the poor performance and scaling of mASPreduce, caused by the limitation of the high-level framework Spark. 
To reach this goal, we opted for a C++ implementation, with the help of the 
Parallel Boost Graph Library (briefly, PBGL) for the distributed graph data structure, and the boost MPI library for the communication stage. Thanks to the latter, we have complete control over the messages sent on the network and the synchronization between the different computational nodes. Since the bad scaling of mASPreduce resides on the communication stage, our optimization starts from that.

The way PBGL distributes the graph is pretty straightforward: vertices are divided between the computational nodes in a Round Robin way, stored in a node list, and each unit keeps track of the edges connected to its local vertices with adjacency lists. The user can choose between different data structures to implement both the node list and the adjacency lists (vector or list), and he can also set a property map, which we use to store vertex characteristics (color, supported, blocked, \dots).\\

Anyway, this naive graph partitioning leads to poor communication performance, caused by the high size of the cut (with cut we mean the set of edges which connect two vertices stored in different computational units).\\
For this reason, we developed a greedy redistribution algorithm, which, for each pair of computational nodes, keeps swapping the two vertices that contribute more to the cut until a fixpoint is reached, i.e., until the swapping operation increases the cut size.

\subsection{Design choices}

The first and most visible change with respect to mASPreduce is a modification of the $\mathcal{RDG}$ structure, which has two noticeable effects: it is more suitable to address the $\mathit{notify\_change}$ implementation of propagation, explained later in this section, and it can considerably reduces the number of edges, at the cost of doubling up the nodes.  
From now on, we refer to such a graph as $\mathcal{RDG}$'.

\begin{definition}[New Rule Dependency Graph: $\mathcal{RDG}$']
Given a logic program $\Pi$, we define the $\mathcal{RDG}$' $\Gamma$ as the graph $(V,E_0,E_1,E_2)$ where
\begin{itemize}
\item
$V = \Pi \: \cup \: atoms(\Pi)$
\item
positive edges  $E_0$: they have the same meaning they had in $\mathcal{RDG}$, but the source node must be an atom $a$ and the destination node a rule $r$:
\[E_0 = \{(a,r) \ | \ a \in atoms(\Pi) \ , \ r \in \Pi, a \in body^+(r) \} \]
\item
negative edges  $E_1$: they have the same meaning they had in $\mathcal{RDG}$, but the source node must be an atom $a$ and the destination node a rule $r$:
\[E_1 = \{(a,r) \ | \ a \in atoms(\Pi) \ , \ r \in \Pi, a \in body^-(r) \} \]
\item
head edges $E_2$: they link each rule with its head (if it has any):
\[E_2 = \{(r,a) \ | \ r \in \Pi \  , \ a \in atoms(\Pi), head(r) = a \} \]
\end{itemize}

\end{definition}

The reason why this graph is more suitable to our algorithm is that we rely only on information local to a node to decide whether the latter is supported or blocked.\\ For instance, a rule $r$ is unblocked if we are sure that in the actual coloring an atom $a$ belonging to $body^-(r)$ does not belong to the answer set, i.e, for all rules $r'$ such that $head(r') = a$, then $r' \in \C_\ominus$. To perform this check without forcing $r$ to query all its neighbors, we could use a counter for each atom in $body^-(r)$ to count how many $r'$ were disabled. Since it is not a good idea to keep variable size data structures inside a node, we opted to use atom nodes, each one with its own single counter.\\
\newline
The other reason to choose this $\mathcal{RDG}$ structure is that it can strongly decrease the number of edges, which is a very good point in a distributed graph: the fewer edges between different computational nodes, the less amount of communication.\\
To explain this property, the reader can imagine a logic program with $n$ rules with the same head $a$, which in turn is present in the body of $m$ rules. To represent $a$'s dependencies, an $\mathcal{RDG}$ would need $n*m$ edges (from each of $n$ rules to each of $m$ rules), while an $\mathcal{RDG}$' can instead obtain the same result with only $n+m$ edges (from each of $n$ rules to the atom node $a$, and from the atom node $a$ to each of $m$ rules).

\medskip

To address performance and reduce communication, a completely different strategy was developed in DASC to implement the propagation operators, as the reader can notice below. 

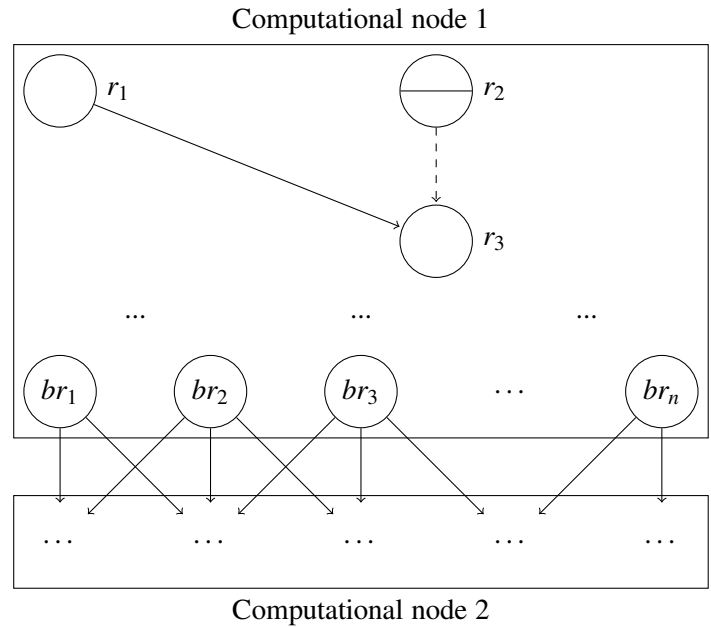
\begin{figure}
\begin{center}
\begin{tikzpicture}
        \node[state, label={right:$r_1$}] at (0,4)  (s1) {};
        \node[state,label={right:$r_2$},circle split] at (5,4)  (s2) {};


        \node[state,label={right:$r_3$}] at (5,2)  (p) {};

         \node[state, draw=none] at (1,1)  (p2) {...};
        \node[state, draw=none] at (4,1)  (p3) {...};
         \node[state, draw=none] at (7,1)  (p4) {...};

        \node[state] at (0,0)  (s4) {$br_1$};
        \node[state] at (2,0)  (s5) {$br_2$};
        \node[state] at (4,0)  (s6) {$br_3$};
        \node[state, draw=none] at (6,0)  (s7) {\dots};
        \node[state] at (8,0)  (s8) {$br_n$};

         \node[state, draw=none] at (0,-2)  (c4) {\dots};
        \node[state, draw=none] at (2,-2)  (c5) {\dots};
        \node[state, draw=none] at (4,-2)  (c6) {\dots};
        \node[state, draw=none] at (6,-2)  (c7) {\dots};
        \node[state, draw=none] at (8,-2)  (c8) {\dots};

        \node [draw=black, label= Computational node 1, fit= (s1) (s2) (p) (s4) (s5) (s6) (s7) (s8) ] {};
		 \node [draw=black, label= {below:Computational node 2}, fit= (c4) (c5) (c6) (c7) (c8) , ] {};

          {end};

        \draw[every loop]
        (s1) edge node {} (p)
        (s2) edge [dashed] node  {} (p)

         (s4) edge node {} (c4)
         (s4) edge node {} (c5)
          (s5) edge node {} (c5)
          (s5) edge node {} (c4)
          (s5) edge node {} (c6)
          (s6) edge node {} (c5)
          (s6) edge node {} (c6)
          (s6) edge node {} (c7)
          (s8) edge node {} (c7)
          (s8) edge node {} (c8);

    \end{tikzpicture}
    \end{center}
\caption{This figure represents the main issue of the MapReduce paradigm: during a propagation involving only nodes owned by a single machine (like $r_1,r_2,r_3$), the border nodes $br_i$ do query their neighbors (owned by computational node 2), generating useless communication.}
\label{fig:comm}
\end{figure}

The MapReduce paradigm has a big downside when dealing with a distributed system. Querying a neighbor stored in another computational node is a very expensive operation, and this situation always happens, even if the considered vertex would never be touched by the actual propagation. Looking at the example in Figure \ref{fig:comm}, we refer to nodes connected to other computational units as border nodes; since MapReduce relies on the fact that each node queries all of its neighbors, this implies that also in the case of a local propagation (like one involving only $r_1$ and $r_3$), which theoretically does not need to send any message on the network, edges connected to border nodes are crossed, causing useless traffic inside the cluster.

To fix the problem, the idea is to develop an algorithm in which only the nodes really affected by the actual propagation (plus their neighbors) are touched: we will refer to this implementation as $\mathit{notify\_change}$ algorithm, since it will be duty of an affected node to notify its neighbors of an eventual change in its coloring state, and not the opposite.

\section{Preliminary results}

\begin{figure}
\begin{lstlisting}
dom(1..n).

sel(X) :- dom(X), not nsel(X).

nsel(X) :- dom(X), not sel(X).

:- sel(X), sel(Y), X != Y.

p(X1,X2,X3,X4,X5,X6) :- 
  sel(X1), sel(X2), sel(X3), sel(X4), sel(X5), sel(X6).
\end{lstlisting}
\caption{Toy example: by increasing $n$ inside $\textit{dom(1..n)}$, the number of generated ground roules grows exponentially}
\label{fig:toy}
\end{figure}

In Figure \ref{fig:testD}  the reader can find a quick comparison between DASC and mASPreduce. We tested a toy example (Figure \ref{fig:toy}) in which, by only changing the domain size of the problem, it is easy to generate ground programs with a high number of ground rules: since our program does not (yet) make use of any heuristics, and we are more interested in how it deals with a huge number of nodes to distribute, this behaviour is useful to test how well a problem scales over both the number of computational nodes and the number of vertices in the RDG.\\
Each DASC test is executed with all possible combinations of distribution options:
\begin{itemize}
\item
Distribution algorithm: it could be either a naive round robin distribution (standard behaviour of $boost$) or a greedy distribution algorithm which tries to minimize cut size;
\item
Number of computational nodes: each problem is tested using different numbers of computational units of the cluster, from 1 to 5.
\end{itemize}
We omit the Clingo (single thread) tables since almost all timings were at most 10ms.

\newcommand\spa{0.95cm}
\begin{figure}
\begin{tabular}{|l|l|p{\spa}|p{\spa}|p{\spa}|p{\spa}|p{\spa}|p{\spa}|p{\spa}|p{\spa}|p{\spa}|p{\spa}|}
\hline
\multirow{2}{*}{Inst} & \multirow{2}{*}{Distr}  & \multicolumn{2}{|c|}{1 cp unit}  &  \multicolumn{2}{|c|}{2  cp unit} & 
 \multicolumn{2}{|c|}{3  cp unit} &  \multicolumn{2}{|c|}{4  cp unit}  &  \multicolumn{2}{|c|}{5  cp unit}  \\ 
 && DASC & MR &  DASC & MR & DASC & MR & DASC & MR & DASC & MR \\
 \hline
\multirow{2}{*}{1} & RR  & 0.003 &  \multirow{2}{*}{56.330} & 0.013  &  
\multirow{2}{*}{42.190} & 0.015 & \multirow{2}{*}{40.160} & 0.015 & 
\multirow{2}{*}{41.177} & 0.017  & \multirow{2}{*}{35.405} \\ 
\cline{2-3}  \cline{5-5} \cline{7-7} \cline{9-9} \cline{11-11}
                  & RD &NR && 0.010 && 0.011  && 0.013  && 0.014  &   \\ \hline
 \multirow{2}{*}{2} & RR & 0.048  & \multirow{2}{*}{95.697 }  & 0.14 & \multirow{2}{*}{64.315}  
& 0.19  & \multirow{2}{*}{61.767}  & 0.16 & \multirow{2}{*}{62.845}  & 0.19 & \multirow{2}{*}{54.144 } \\ 
\cline{2-3}  \cline{5-5} \cline{7-7} \cline{9-9} \cline{11-11}
                  & RD  &NR && 0.184 && 0.151 && 0.166  &&  0.172 &    \\ \hline
\multirow{2}{*}{3} & RR & 0.36  
&\multirow{2}{*}{150.82 } & 1.11  & \multirow{2}{*}{88.043} & 1.42  & \multirow{2}{*}{89.145 } &
 1.36 &\multirow{2}{*}{89.695} &  1.33 & \multirow{2}{*}{78.178 } \\ 
\cline{2-3}  \cline{5-5} \cline{7-7} \cline{9-9} \cline{11-11}
                  & RD  &NR && 1.226 && 1.393 && 1.115  &&  1.232  &  \\ \hline
\multirow{2}{*}{4} & RR & 1.83  &\multirow{2}{*}{SE} & 6.23  &\multirow{2}{*}{SE} & 
6.18  &\multirow{2}{*}{SE} & 6.26 &\multirow{2}{*}{SE} & 5.98 &\multirow{2}{*}{SE}  \\ 
\cline{2-3}  \cline{5-5} \cline{7-7} \cline{9-9} \cline{11-11}
                  & RD  &NR && 6.118 && 6.401 && 6.226  &&  5.256 &    \\ \hline
\multirow{2}{*}{5} & RR & 7.03  &\multirow{2}{*}{to} & 22.84 &\multirow{2}{*}{to} &
 23.86 &\multirow{2}{*}{to} & 33.60 &\multirow{2}{*}{to} & 24.21 &\multirow{2}{*}{to} \\ 
\cline{2-3}  \cline{5-5} \cline{7-7} \cline{9-9} \cline{11-11}
                  & RD  &NR && 22.513 && 20.765 && 20.881  &&  18.511  &   \\ \hline
\multirow{2}{*}{6} & RR & 21.99  & \multirow{2}{*}{to} & 71.09  &\multirow{2}{*}{to} & 81.55  &\multirow{2}{*}{to} & 69.76  &\multirow{2}{*}{to} & 65.83 &\multirow{2}{*}{to}\\ 
\cline{2-3}  \cline{5-5} \cline{7-7} \cline{9-9} \cline{11-11}
                  & RD  &NR && 71.07 && 83.60 &&  66.43 &&  68.20  &   \\ \hline
\multirow{2}{*}{7} & RR & 58.90  & \multirow{2}{*}{to} &188.85  &  \multirow{2}{*}{to} &185.45 &\multirow{2}{*}{to} & 212.69  &\multirow{2}{*}{to} &  220.73 & \multirow{2}{*}{to} \\ 
\cline{2-3}  \cline{5-5} \cline{7-7} \cline{9-9} \cline{11-11}
                  & RD  &NR   && 185.46  && 195.41   && 182.33   &&  191.27 &   \\ 
\hline
\end{tabular}
\caption{Comparison between DASC and mASPreduce (MR) on a set of benchmarks.
For DASC, two distribution options are tested: round robin (RR) and greedy redistribution (RD).
NR means ``not relevant'' (we can not test a redistribution modality with 1 cp unit), SE ``Spark Error'', and ``to'' timeout. Tests involve from 1 to 5 computation units.}
\label{fig:testD}
\end{figure}

It is clear that there is a huge performance gap between both mASPreduce with DASC, and the latter with Clingo.\\
Moreover, also DASC seems to do not scale very well, how the reader can notice by the fact that the tests with only one computational node are the ones which behaviour better in all the cases.
The reason for such a bad scaling lies on the communication phase: apparently, the improvement for having a propagation process parallelized between more nodes is not enough to compensate the performance degradation caused by the communication between those nodes. There are other reasons to still use this distributed approach, like the fact we can theoretically handle programs too big to be contained into a single machine. Yet for the moment, such problems are too complicated to be solved by our tool within acceptable timings, but this will change in the future, through the implementation of various heuristics.\\
Moreover, a big part of the guilt of this behaviour is due to the initial round robin distribution from which the redistribution algorithm starts. In fact, we noticed a very few iterations were made during the graph repartition, resulting in a node distribution too similar to the initial one. We stongly believe that, by addressing this problem, we would obtain a far better scaling.\\

\section{Conclusion and Future Work}
DASC represents a step forward in building a tool capable of exploiting distributed system resources in order to manage huge-size programs, thanks also to the fact that it is capable of solving domains that do not work properly with mASPreduce, like the ones from the 2015 ASP competition. Yet, we are still far from achieving the goal of large problems handling, and a lot of work has to be done to make our tool competitive with state-of-the-art solvers. Heuristics implementation would probably be the main task to perform in order to close the gap with them. At that point, DASC could be used to handle programs too big for single machine solvers.\\
Moreover, it is evident from the performance difference between our tool and mASPreduce that lowering the level of implementation paid off,  together with the development of a different propagation technique, the so-called \textit{notify\_change} approach.\\
To confirm C++ boost improvement with respect to SPARK, in the instances in which Clingo exceeds 10ms, the minimum machine time, DASC is about 500 times slower; STRASP instead, the distributed grounder developed by Pietro Totis in his thesis using SPARK \cite{totisthesis}, capable of solving stratified programs (non-definite programs solvable without non-determinism in polynomial time), is circa 2000 times slower than Clingo.\\
\newline
Finally, we present below the roadmap for DASC:
\begin{itemize}
\item
improving the initial distribution or changing the redistribution algorithm with a more sophisticated one;
\item
implementing multithreading. We identified two kinds of parallelization we can apply, and the program is already prepared for the first one:
\begin{itemize}
\item
task parallelization: in the actual state, each time a node receives a message with a distribution task to execute, this is stored inside a stack. Since we noticed during the testing that this stack can reach really huge size, a good idea would be to process these tasks in parallel, by distributing them between more threads. Of course this kind of parallelization would work only in a distributed system;
\item
 local parallelization: to exploit multithreading also when working in a local machine we can do the following: each time we need to propagate some information from a vertex with many outgoing edges, we can cross each edge with a different thread, in order to parallelize the various propagation branches.
\end{itemize}
Obviously multithreading introduces some issues, like the fact we need to avoid that more threads/propagation branches read or write from the same vertex at the same time, because one of them could retrieve invalid information, but Boost provides us all the tools to address this problem, like mutex access;
\item
implementing heuristics, like the strong techniques used by Clingo, namely clause learning and backjumping.

\end{itemize}

\bibliography{mybib.bib}

\end{document}